\documentclass[aps,prd,twocolumn,nofootinbib]{revtex4}

\usepackage{amsmath,amssymb,amsfonts,epsfig,color}
\usepackage[retainorgcmds]{IEEEtrantools}
\usepackage{subfigure}

\newcommand{\be}{\begin{equation}}
\newcommand{\ee}{\end{equation}}
\newcommand{\ben}{\begin{eqnarray}\displaystyle}
\newcommand{\een}{\end{eqnarray}}

\newcommand{\bhat}[1]{\hat{\mathbf{#1}}}

\begin{document}

\title{Parity in the CMB: Space Oddity}
\author{Assaf Ben-David}
\email{bd.assaf@gmail.com}
\author{Ely D. Kovetz}
\email{elykovetz@gmail.com}
\author{Nissan Itzhaki}
\email{nitzhaki@post.tau.ac.il}
\affiliation{Raymond and Beverly Sackler Faculty of Exact Sciences,
School of Physics and Astronomy, Tel-Aviv University, Ramat-Aviv, 69978,
Israel}

\begin{abstract}

We search for a direction in the sky that exhibits parity symmetry under reflections through a plane. We use the natural estimator, which compares the power in even and odd $\ell+m$ multipoles, and apply minimal blind masking of outliers to the ILC map in order to avoid large errors in the reconstruction of multipoles. The multipoles of the cut sky are calculated both naively and by using the covariance inversion method and we estimate the significance of our results using $\Lambda$CDM simulations. Focusing on low multipoles, $2\leq \ell \leq \ell_{\max}$ with 
$\ell_{\max}=5,6$ or even 7, we find two perpendicular directions of even and odd parity in the map. While the even parity direction does not appear significant, the odd direction is quite significant -- at least a $3.6\sigma$ effect.

\end{abstract}

\maketitle

\section{Introduction}

Since the release of the COBE data and especially after the release of the WMAP data many groups argued that there are unexpected features at the largest possible scales (see \cite{WMAP7,Copi:2010na} and references within), typically at $2\leq \ell \leq 5$, (for a partial list, see \cite{Tegmark, Copi, Land1}). A real anomaly at the largest scale, if exists, could be explained by pre-inflationary physics. Thus having a better understanding of these potential anomalies might tell us something about the universe before inflation.

Cosmic variance and foregrounds make it harder to estimate the significance of the various claims in the literature. Moreover, a standard argument against most of these potential anomalies is the lack of motivation to look for them to begin with. Or as has been often remarked (see e.g. \cite{WMAP7}), when applying enough statistical estimators to a given data set with random behavior, ``anomalies" are bound to appear. Estimating their significance, one tends to ignore this fact and focus on the anomalous finding alone.

With this concern in mind we focus in this paper on a simple question that needs no justification to be asked: does the CMB at the largest scales behave as expected with respect to a mirror parity transformation? This is not the first time this question was asked. The authors of \cite{Land1} studied the parity transformation of the CMB with respect to the direction of the alignment of the $\ell=2,3$ multipoles \cite{Land2}. Recently, \cite{NaselskyKim} studied the parity of the CMB under reflections through the galactic plane. Here we study the parity symmetry of the CMB data with respect to all directions in the sky and check for anomalous parity directions.

The paper is organized as follows. In section \ref{sec:method} we define a natural estimator for parity symmetry with respect to reflections through a plane normal to some direction $\bhat{n}$ and present the masking scheme we use. In section \ref{sec:results} we generate parity maps for the full and cut WMAP seven-year {\it Internal Linear Combination} (ILC) map that show the parity score as a function of $\bhat{n}$. We find a parity-even direction that is related to the alignment of the $\ell=2,3$ multipoles and is not significant. More interestingly there is a significant (about $3.6\sigma$) parity-odd direction. In section \ref{sec:conclusions} we conclude.

\section{Method}
\label{sec:method}

\subsection{Full Sky Parity Estimator}

We would like to check the CMB temperature fluctuations map for parity symmetry with respect to reflections through a plane, i.e. $\bhat{r}\to\bhat{r}-2\left(\bhat{r}\cdot\bhat{n}\right)\bhat{n}$, where $\bhat{n}$ is the normal to the plane.
Using the full sky it is most convenient to work with the spherical harmonics $Y_{\ell m}(\bhat{n})$ which transform as $(-1)^{\ell+m}$ under this operator. Hence a natural parity estimator is
\be
\tilde{S}(\bhat{n}) = \sum^{\ell_{\max}}_{\ell=2}\sum^{\ell}_{m=-\ell} (-1)^{\ell+m}\frac{\left| a_{\ell m}(\bhat{n}) \right|^2}{\hat{C}_{\ell}}\, , \label{ParityScore}
\ee
where  $\bhat{n}$ is the $z$-axis used in the harmonic expansion and $ \hat{C}_\ell = \frac{1}{2\ell +1}\sum_m \left| a_{\ell m}\right|^2$.

The more $\tilde{S}(\bhat{n})$ is positive (negative) the more the temperature map is parity-even (odd) in the $\bhat{n}$ direction. Since the ensemble average does not vanish, $\langle \tilde{S} \rangle = \ell_{\max}-1$,  it is convenient to redefine the parity estimator as
\be
S(\bhat{n})=\tilde{S}(\bhat{n}) - (\ell_{\max}-1).
\ee
We shall use this estimator in the next section.

This simple parity estimator is useful when dealing with full sky data. When working with actual  data which suffers from foreground noise, one needs to mask some parts of the sky. The generalization of this parity estimator to this case is the subject of the rest of this section.

\subsection{Masking Scheme}

Naturally, we would like the results to be insensitive to foregrounds. The cleanest approach would be to use the full WMAP KQ75 or KQ85 masks. Unfortunately, these masks cover roughly $30\%$ and $20\%$ of the sky, respectively. Such extremely large masks induce significant errors in the calculations of the multipole coefficients regardless of the reconstruction method used to calculate them and thus render the parity estimator unreliable.

In an attempt to overcome this issue, we would like to define a mask that is as small as possible (and so it can be used to calculate multipole coefficients) but is still efficient in removing local outlying regions and so enables to test {\it large scale physics}. We emphasize that the mask defined below is by no means an alternative to the standard masks  used to extract small scale properties of the CMB.

\begin{figure}
\centering
\includegraphics[width=\linewidth]{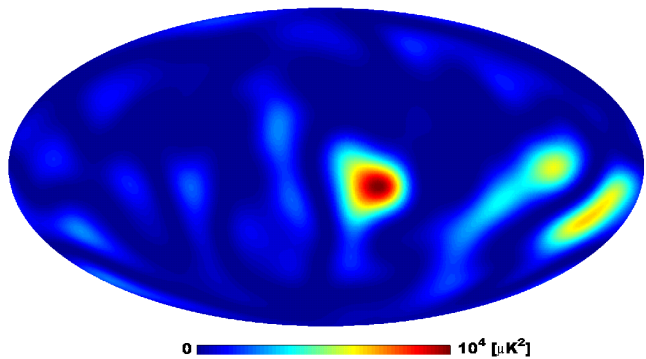}
\caption{The squared smoothed ILC map. The prominent regions lie close to the galactic plane, or inside it.}
\label{dT2}
\end{figure}

To pinpoint the regions that induce noise at large scales we examine the squared temperature map after smoothing it by $20^{\circ}$, as shown for the ILC in Fig.~\ref{dT2}.
This is motivated by \cite{WMAP7}, where this map of anisotropy power was used to manually define localized regions and check the sensitivity of the alignment of $\ell=2,3$ to their removal. Here we simply automate this procedure to choose the most outlying local regions in the map. Not surprisingly, it is readily seen that the most intense areas in the map lie in small patches around the galactic plane. One option would be to mask out only the intense regions that are inside the KQ75 mask.  This option, however, is problematic when estimating the significance of the result vs. random simulations of $\Lambda$CDM. The reason is that such an estimation requires an automatic procedure for defining the masked regions. But the overlap between the most intense regions of a random map and the KQ75 mask is very different than the overlap between the most intense regions of the ILC and the KQ75 mask.

\begin{figure}
\centering
\includegraphics[width=\linewidth]{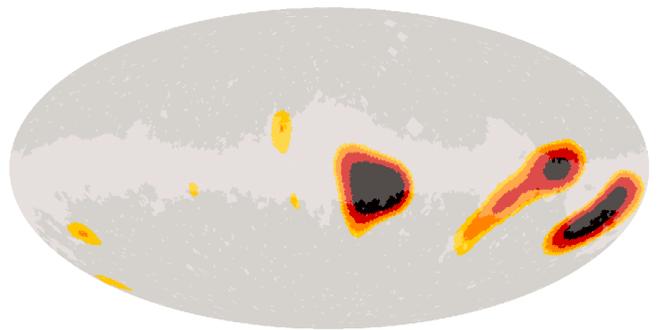}
\caption{Masks that cover the most powerful pixels of the squared, smoothed ILC map. The masks cover $2.5\%$ (\emph{black}), $5\%$ (\emph{red}), $7.5\%$ (\emph{orange}) and $10\%$ (\emph{yellow}) of the sky. The KQ75 galactic mask is also shown (in light gray).}
\label{ILCMasks}
\end{figure}

An alternative option, which we do follow,  is to mask out a fixed total area of deviating pixels. For our analysis we examine the masking of the most intense areas from $1\%$ to $20\%$ of the sky. For the ILC map, masks covering $2.5\%,5\%,7.5\%$ and $10\%$ of the sky are shown in Fig.~\ref{ILCMasks}, together with the galactic KQ75 mask, for reference.
Most of these masked regions lie inside or near the KQ75 galactic mask.\footnote{These regions overlap with the regions identified in \cite{Hajian} as the ones containing most of the large scale angular power in the ILC behind the KQ75 galactic mask.} To quantify this, we calculate for a given masking area, $A$, the area that lies within the KQ75 mask, $A^*$, and compare this to random $\Lambda$CDM simulations. We find that for $5\%\le A\le 15\%$, only $\sim\!0.6\%$ of the simulations have a greater $A^*$ than that of the ILC map. This provides some justification for our choice of masking these local regions around the galactic plane in the case of current WMAP data.

Generally speaking there are two ways to calculate the multipole coefficients $a_{\ell m}$ on a masked sky (both break down if the mask is too large). The naive method is to zero the region inside the mask and take a spherical harmonic transform. A more precise method is the `covariance inversion method' used in the analyses of \cite{OliveiraCosta, Efstathiou, Aurich} which we discuss next.

The relation between the temperature map and the spherical harmonic coefficients is
\be
\mathbf{x} = \mathbf{Ya}+\mathbf{n}\, ,
\ee
where $\mathbf{x}$ is a vector of the temperature in the masked sky, $\mathbf{Y}$ is a matrix of the spherical harmonics evaluated on each direction ($\mathbf{Y}_{ij}=Y_{\ell_j m_j}\left(\bhat{r}_i\right)$), $\mathbf{a}$ is the coefficients vector and $\mathbf{n}$ represents the noise in the data. Assuming Gaussianity and statistical isotropy and ignoring the detector noise, the familiar solution for the reconstructed coefficients $\bhat{a}$ is
\be
\bhat{a} = \left(\mathbf{Y}^\dagger \mathbf{C}^{-1} \mathbf{Y}\right)^{-1}\mathbf{Y}^\dagger\mathbf{C}^{-1}\mathbf{x}\, .
\ee
The square matrix $\mathbf{C}$ is constructed from a given a power spectrum $C_{\ell}$ as
\be
\mathbf{C}_{ij}=\sum_{\ell=\ell_{\max}+1}^L \frac{2\ell+1}{4\pi}P_\ell\left(\bhat{r}_i\cdot\bhat{r}_j\right)C_\ell\, , \label{Cij}
\ee
where $P_\ell$ is the Legendre polynomial. The sum in (\ref{Cij}) runs over those multipoles that are not included in the coefficients vector. In this manner, this method compensates for the contribution of higher-$\ell$ multipoles that leak into the reconstructed coefficients since the spherical harmonics are no longer orthogonal on the cut sky. For the analysis in this work we use the seven year WMAP power spectrum and take $L=196$, roughly corresponding to a scale of $1^\circ$.

\subsection{Cut Sky Parity Estimator}

Using the masking scheme discussed above, we replace our full sky estimator $S(\bhat{n})$ with a corresponding cut sky version
\begin{IEEEeqnarray}{l}
S(\bhat{n}, A)   = \nonumber \\ 
\sum^{\ell_{\max}}_{\ell=2}\sum^{\ell}_{m=-\ell} (-1)^{\ell+m}\frac{\left| \hat{a}_{\ell m}(\bhat{n},A) \right|^2}{\hat{C}_{\ell}(A)}  -(\ell_{\max}-1)\, ,
\label{ParityScoreCut}
\end{IEEEeqnarray}
where $\hat{a}_{\ell m}(\bhat{n},A)$ and $\hat{C}_{\ell}(A)$ are reconstructed after masking the area $A$ of the most intense pixels in the map.

If there is a direction in the sky, $\bhat{n}_0$, which is parity-even (odd) in an anomalous  way then we expect $S(\bhat{n}_0, A)$ to increase (decrease) significantly  as we increase $A$, since more and more galactic noise is removed. Eventually, as we keep on increasing $A$, the reconstruction noise grows and at some point it is bound to dominate and lower (raise) $S(\bhat{n}_0, A)$. In the next section we check if there are such  directions in the WMAP data. We will mainly use the covariance inversion method. However, since the naive method is much faster to compute and does not rely on a particular power spectrum, we shall use it as well, for comparison.

\section{Results}
\label{sec:results}

\begin{figure}
\centering
\includegraphics[width=\linewidth]{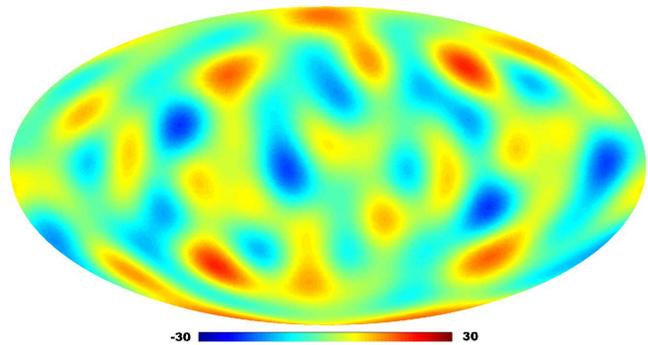}
\caption{The parity estimator calculated on the full sky  7-year WMAP ILC map, with $\ell_{\max}=5$. The maximum lies in the direction of the ``axis-of-evil". The score map is at HEALPix $N_\text{side} = 64$ resolution, corresponding to $1^{\circ}$.}
\label{fullskyscore}
\end{figure}

We begin by calculating the parity score on the full sky WMAP seven-year ILC map degraded to HEALPix resolution $N_\text{side} = 8$ after removing the Doppler-Quadrupole contribution \cite{Kamionkowski}.
The result for $\ell_{\max}=5$ (motivated by \cite{Copi,Land2}) is shown in Fig.~\ref{fullskyscore}.
As can be seen, the maximum lies at $(l,b)=(260^{\circ},48^{\circ})$ which is exactly the direction found in \cite{Tegmark} of the renowned alignment of the quadrupole and octupole. Since the alignment of the multipoles is directly related to their planarity (corresponding to large $m=\pm\ell$ modes), this can be seen as a special case of an even parity symmetry (large even $\ell+m$ modes) \cite{Land2}.
On the full sky, this direction does not seem significant.

We also notice a minimum in the map at $(l,b)=(266^{\circ},-19^{\circ})$, signifying odd parity. However, this minimum as well does not appear to be significant on the full sky map.

As discussed in \cite{WMAP7}, the alignment of the quadrupole and octupole is extremely sensitive to the masking of small patches near the galactic plane and even with small percentages of the sky removed, the alignment is destroyed. To check the stability of the parity results, we use the masking scheme described above and check the behavior of our estimator on the cut sky. When degrading the cut sky map to $N_\text{side} = 8$, we follow the prescription of \cite{Aurich} whereby each downgraded pixel is calculated by averaging only the corresponding unmasked original pixels in the higher resolution, unless more than half of them are inside the higher resolution mask, in which case the downgraded pixel is masked as well.

\begin{figure*}
\centering
\subfigure[]{\includegraphics[width=0.45\linewidth]{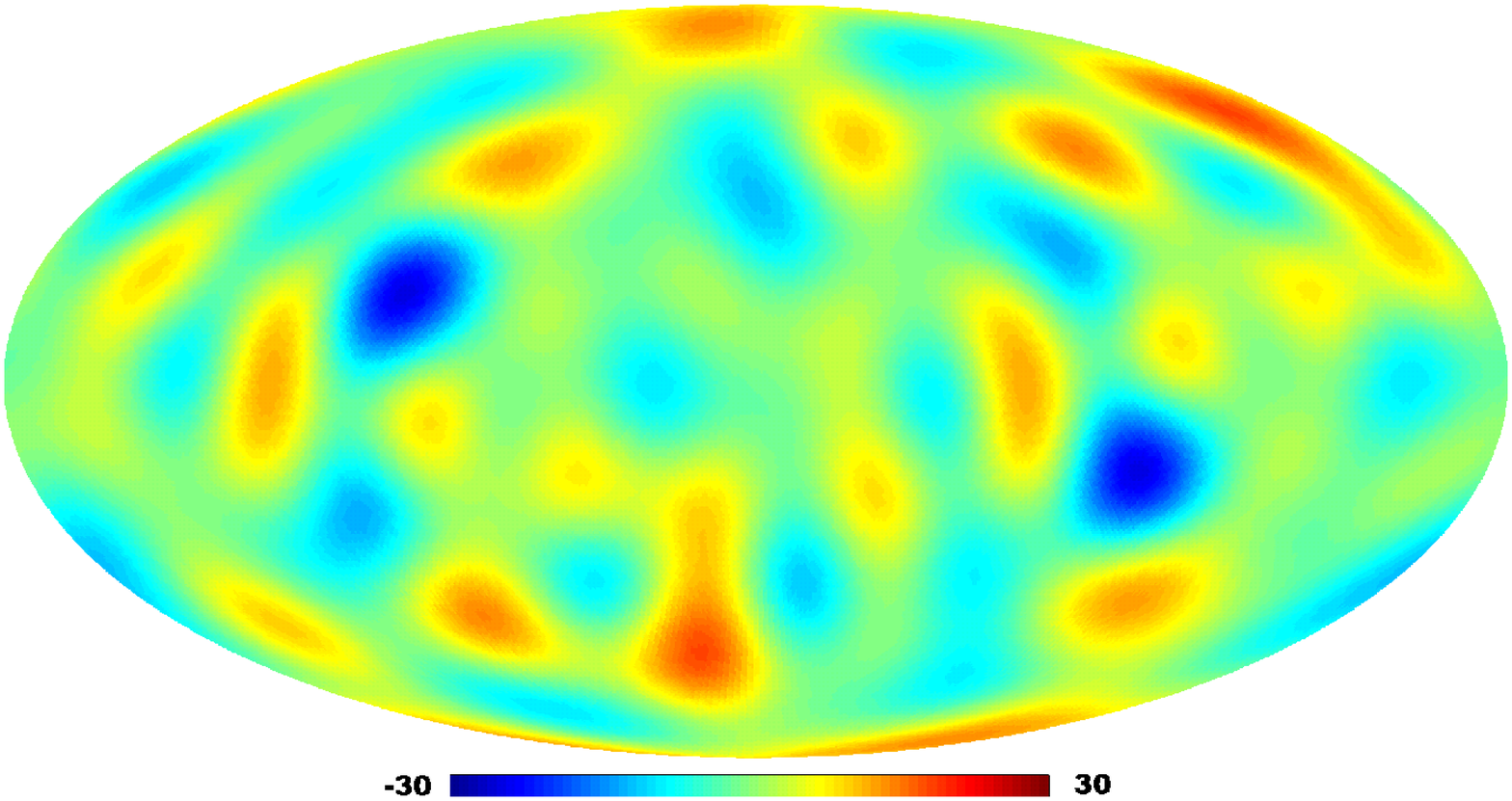}}
\subfigure[]{\includegraphics[width=0.45\linewidth]{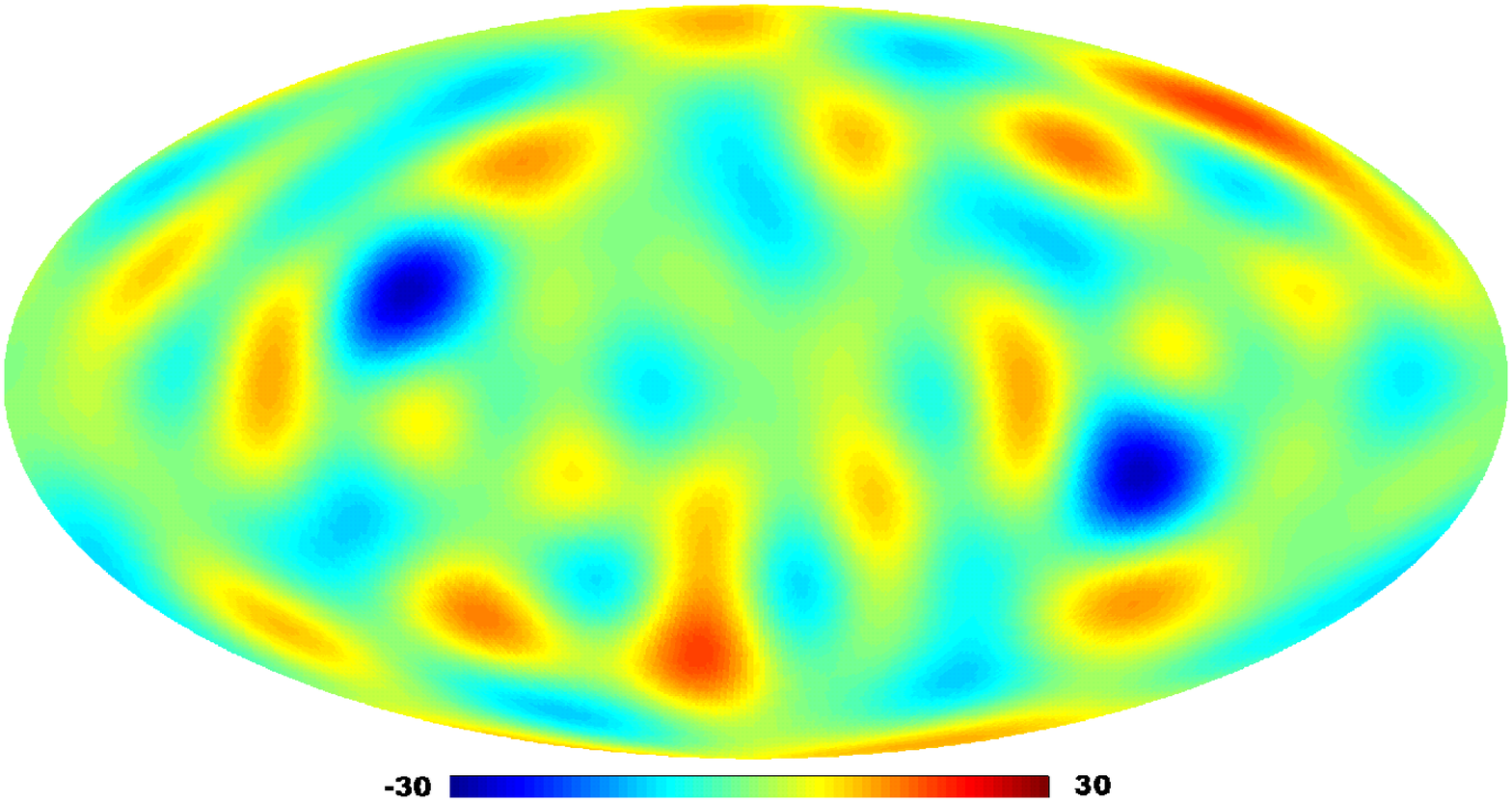}}
\caption{The parity estimator calculated on the ILC map with part of the sky masked as described in the text. The masks used in panels (a) and (b) cover $10\%$ and $12\%$ of the sky, respectively. The multipole coefficients are reconstructed using the covariance inversion method.}
\label{maskedskyscore}
\end{figure*}

The results using two representative masks covering the most powerful $10\%$ and $12\%$ of the sky (both of which, as any masking above $2.5\%$, eliminate the alignment) are shown in Fig.~\ref{maskedskyscore}.
Comparing the results to the full sky map (Fig.~\ref{fullskyscore}), we can see that the location of the maximum has shifted by almost $40^{\circ}$, to $(l,b)=(198^{\circ},55^{\circ})$. It now stands out relative to the other positive peaks, but it does not appear significant.

The result for the minimum of the map (which signifies maximal odd parity) is more compelling. Its location has shifted by no more than a few degrees when applying the masks and remains stable under further cuts (including the full KQ85 or KQ75 masks). Furthermore, it now appears much more significant, as we discuss below.

It is worthwhile to mention that using the cut sky, the resulting even and odd extrema lie at a distance of $90^{\circ}\pm1^{\circ}$ from each other.

The final step in our analysis is the estimation of the significance of our findings. We have seen that on the masked ILC map, we get distinct even and odd parity directions. Noticeably, these directions in the score map, apart from having a high absolute value, also stand out relative to the rest of the map.
To quantify this, we take the maximum and minimum of the score map,
\be
S_{+}(A)=\max_{\bhat{n}}S\left(\bhat{n},A\right)\, , \quad
S_{-}(A)=\min_{\bhat{n}}S\left(\bhat{n},A\right)
\ee
and define
\be
\bar{S}_{\pm}(A)=\left|\frac{S_{\pm}(A)-\mu(A)}{\sigma(A)}\right|\, ,
\ee
where $\mu(A)$ and $\sigma(A)$ are the mean and standard deviation of the score map, respectively. We can now compare these standardized scores to the values of $\bar{S}_{\pm}(A)$ calculated for random simulations.

We generate random maps using the seven year WMAP power spectrum (at a resolution corresponding to the $1^{\circ}$ resolution of the ILC) and degrade them to $N_\text{side}=8$ before calculating the score. In order to compare with the results on the masked ILC, we apply the same masking procedure described above to the random maps.

\begin{figure*}
\centering
\subfigure[]{\label{scoreMax}\includegraphics[width=0.42\linewidth]{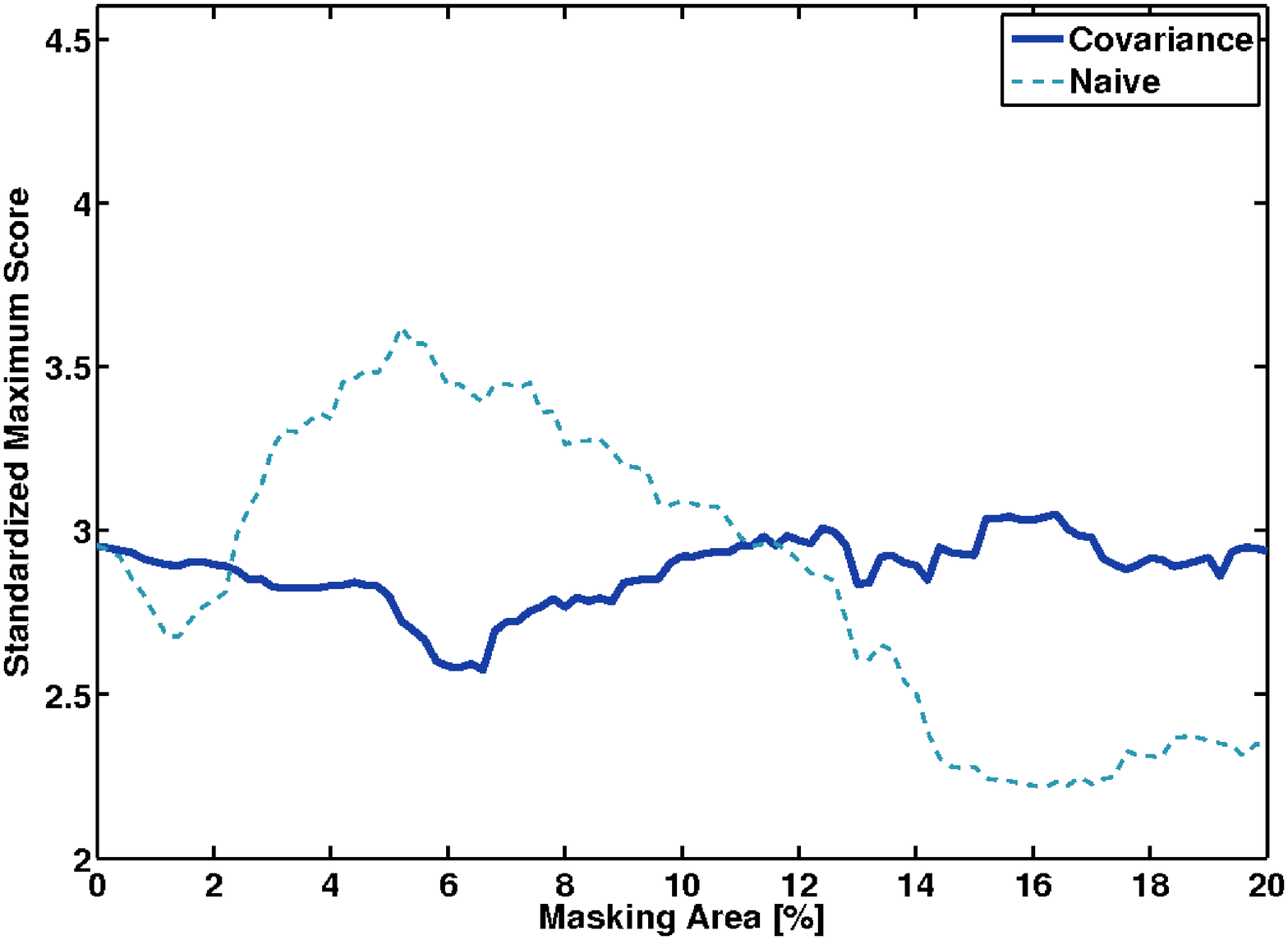}}
\hspace{0.05\textwidth}
\subfigure[]{\label{scoreMin}\includegraphics[width=0.42\linewidth]{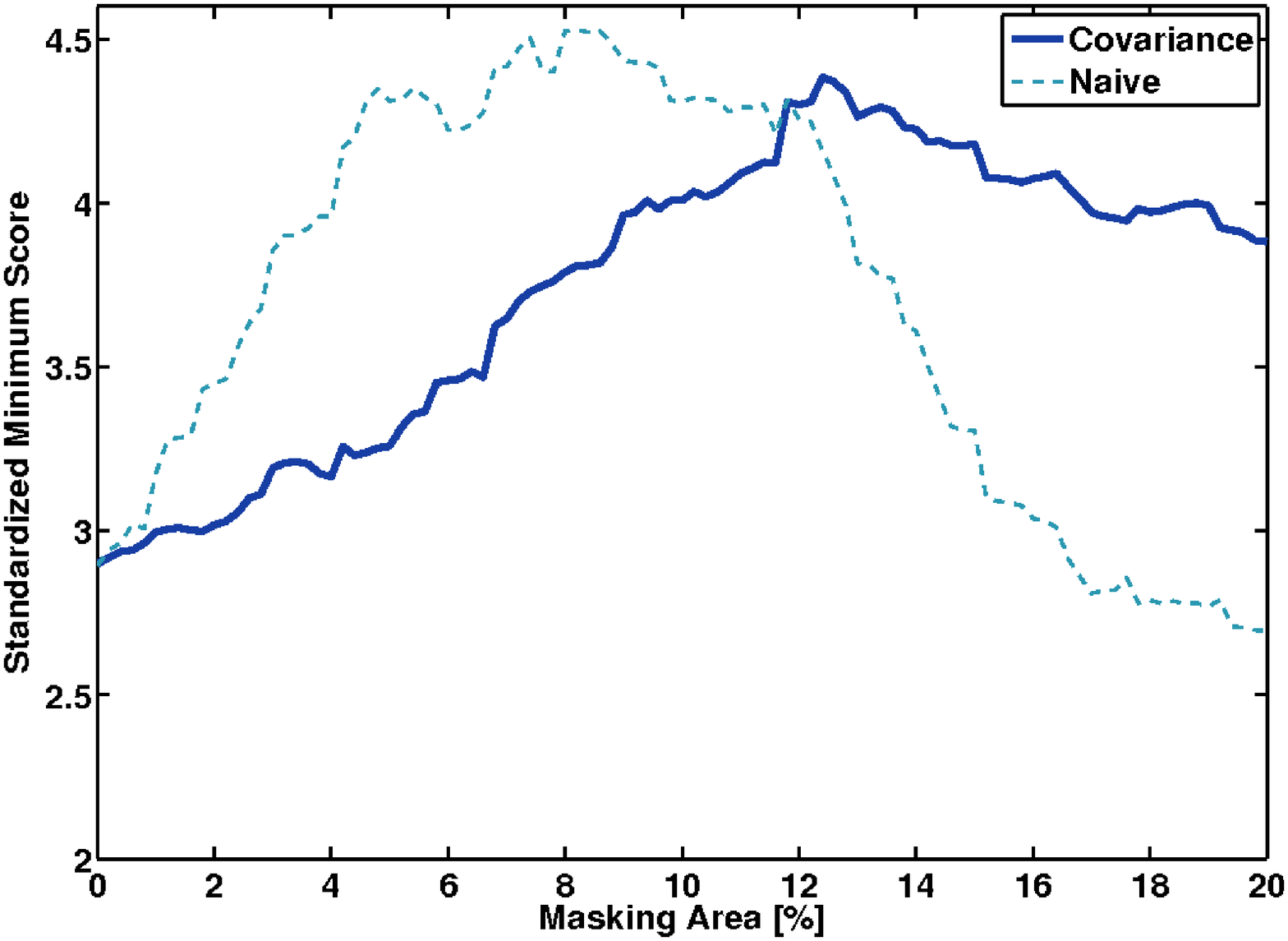}}
\caption{The scores $\bar{S}_{+}$ (a) and $\bar{S}_{-}$ (b) as a function of the masking area, $A$, using the covariance inversion method. The naive result is also plotted, for comparison.}
\label{ScoreFigures}
\end{figure*}

We find that the even parity direction $(l,b)=(198^{\circ},55^{\circ})$ is the maximum of the score map for masking areas larger than $7\%$. However, it does not appear significant. This can be seen from Fig~\ref{scoreMax}. As discussed above if there is a direction in the sky which is parity-even in a significant way then $\bar{S}_{+}(A)$ is expected to increase considerably and consistently as we increase $A$ and eventually decrease as we keep on increasing $A$. This is not the case in Fig~\ref{scoreMax}. Indeed, if we compare the typical value of $\bar{S}_{+}(A)$ of $\sim3$ to random simulations, $12\%$ get a higher score.

The odd parity direction $(l,b)=(264^{\circ},-18^{\circ})$, however, is significant. As can be seen in Fig.~\ref{scoreMin}, when using the covariance inversion method we see that $\bar{S}_{-}(A)$ behaves exactly as expected from an anomalous direction: it increases as we increase $A$ until $A$ is too large and the reconstruction noise takes over. Compared to random simulations, the significance of the peak score at $A=12\%$ is $3.6\sigma$, as $0.03\%$ of $500,\!000$ randoms get a higher score.

When using the naive reconstruction method, we see that the score $\bar{S}_{-}(A)$ follows the same behavior and reaches roughly the same maximum value as the covariance inversion method score. As expected, it peaks at lower masking areas. The reason for this is that the more robust covariance reconstruction compensates for the removed pixels using the given $\Lambda$CDM correlations, while the naive method completely ignores them. Therefore, in order to fully remove the influence of noisy regions, more masking is required when using the covariance inversion method. The discrepancy between the two methods supports the choice of the robust method for estimating the significance of the results, which we have done using the small masks described above.

It is important to note that the odd parity direction remains significant not only for the choice of $\ell_{\max}=5$, but for 6 and 7 as well. In fact, for $\ell_{\max}=6$ it reaches as high as $4.3\sigma$. For larger values of $\ell_{\max}$ the significance starts to drop. However, this is not surprising since the errors in the reconstruction method grow quickly with $\ell$ \cite{OliveiraCosta,Efstathiou,Aurich}. Moreover, as discussed below, if the origin of the anomalous parity behavior is pre-inflationary physics, then it is expected to drop as we increase $\ell$.

\begin{figure}
\centering
\includegraphics[width=0.9\linewidth]{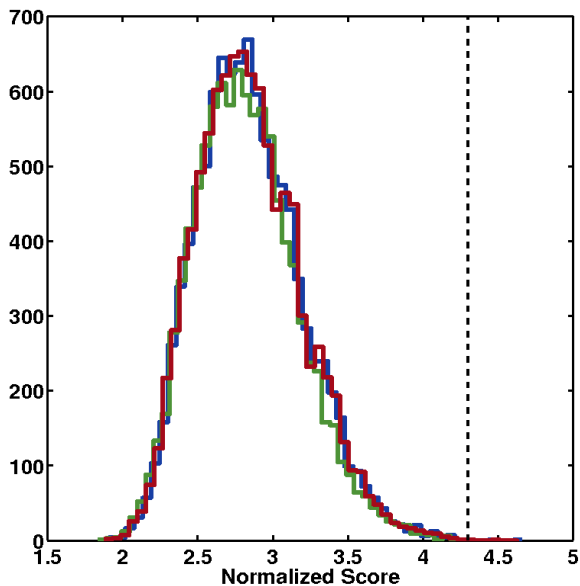}
\caption{Histograms for the score $\bar{S}_{-}$ calculated on $10^{4}$ random realizations, without masking and with $N_\text{side}=8$ (\emph{blue}), with masking of $12\%$ and $N_\text{side}=8$ (\emph{green}) and without masking and with $N_\text{side}=16$ (\emph{red}). The dashed vertical line is at the score value of 4.3, attained by ILC map.}
\label{ScoreHist}
\end{figure}

One might worry that our masking scheme, the reconstruction methods or even our choice of degrading the maps to HEALPix resolution $N_\text{side}=8$ before calculating the score induce bias on the parity score that renders the parity-odd direction so significant. To check the effect on the score and verify that it should not change our conclusions, we plot in Fig.~\ref{ScoreHist} histograms for the score $\bar{S}_{-}$ based on $10^{4}$ random simulations. On each of these random maps three versions of the score were calculated: without masking and with $N_\text{side}=8$, with masking of $12\%$ and $N_\text{side}=8$ and finally without masking but with $N_\text{side}=16$. 
As can be seen, all three histograms are virtually identical. We therefore conclude that the masking, as well as the choice of HEALPix resolution, do not introduce meaningful bias to the score, and do not alter our result. In addition, very recently \cite{Feeney} discussed the covariance inversion method for reconstructing the spherical multipoles and made different choices in smoothing the initial sky map and in the parameters for degrading the map and cutting off the higher multipoles. We verified that the effect of such different choices on our results presented here is negligible as well. 

\begin{figure*}
\centering
\subfigure[]{\includegraphics[width=0.45\linewidth]{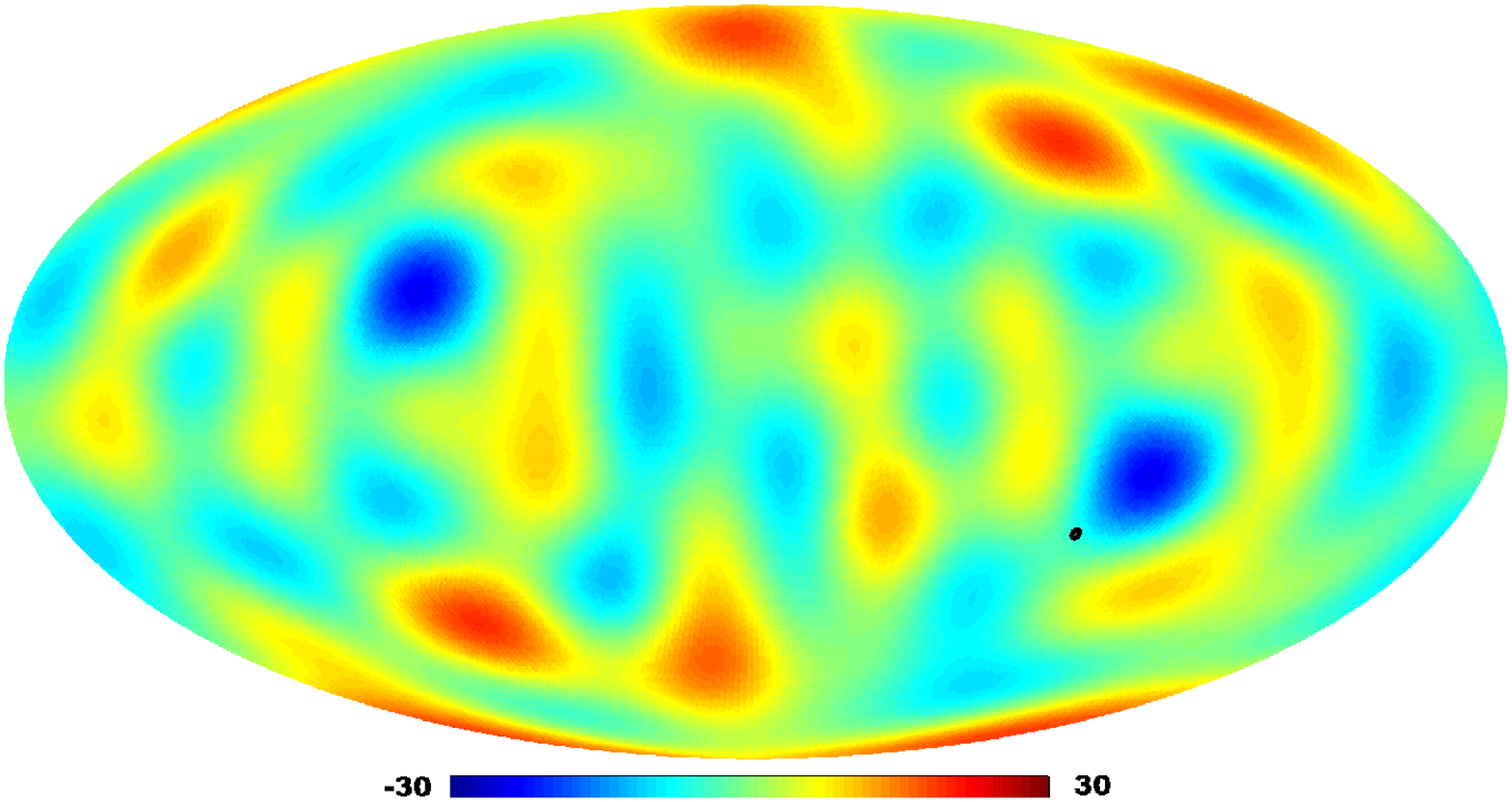}}
\subfigure[]{\includegraphics[width=0.45\linewidth]{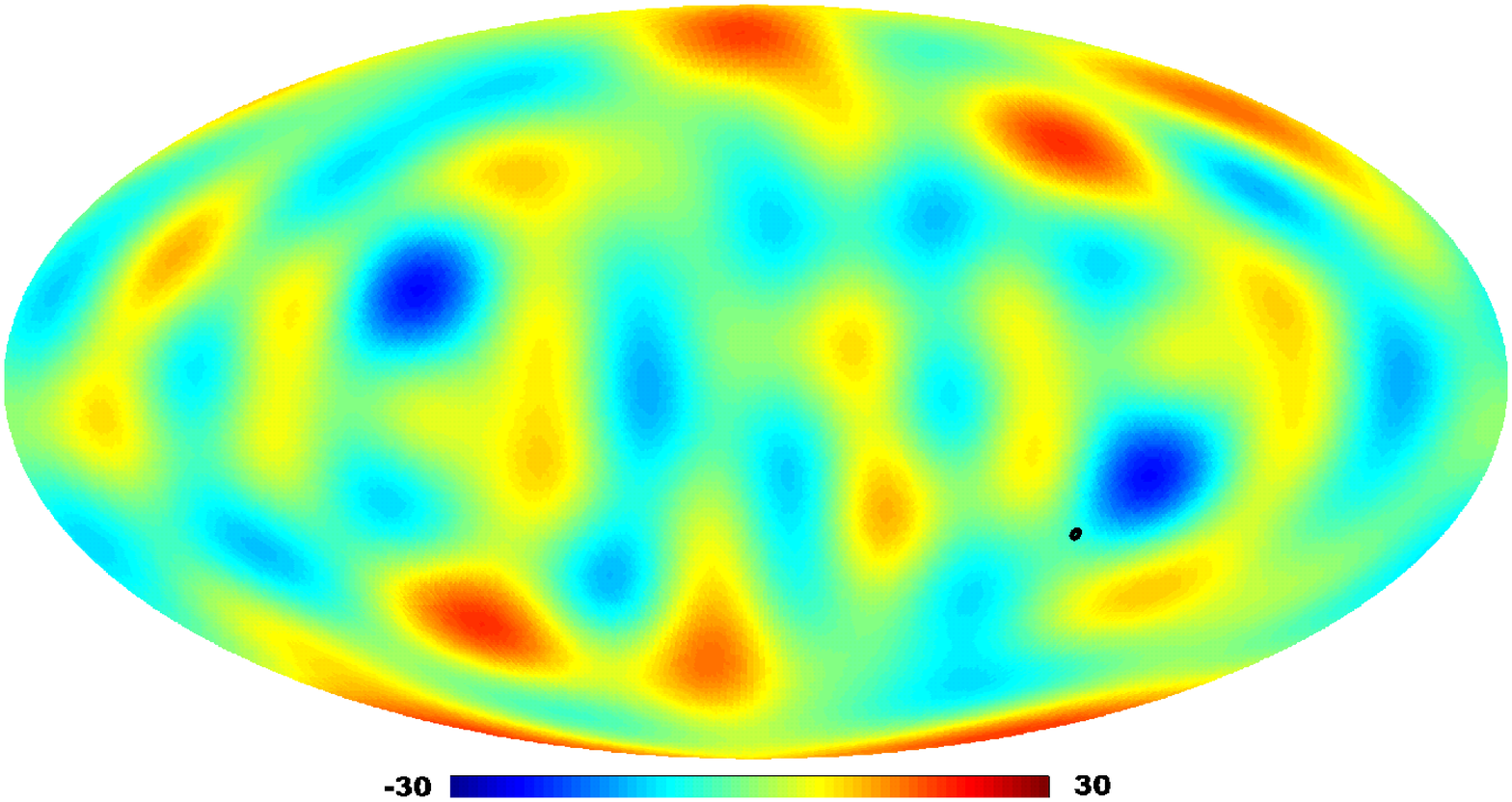}}
\caption{The parity estimator calculated with the covariance inversion method on the V and W frequency band maps with the KQ85 galactic mask applied is shown in panels (a) and (b), respectively. We also mark with a black dot the location of the ecliptic pole, at $(l,b)=(276^{\circ},-30^{\circ})$, where the parity score is neutral.}
\label{VWScore}
\end{figure*}

Finally, to support the possibility of a cosmological origin for our findings, we check our cut sky estimator directly on the WMAP V and W frequency band maps as well. Since in these maps the galactic foregrounds are not removed as in the ILC map, we cannot use the masking procedure described above. Instead, we use the KQ85 galactic mask (covering $\sim\!20\%$ of the sky) for the reconstruction of harmonic coefficients, potentially introducing significant reconstruction errors.
Nevertheless, as is evident in Fig.~\ref{VWScore}, both bands show very similar odd and even parity signals, at the same locations.

\section{Conclusions and Discussion}
\label{sec:conclusions}

We have found strong evidence for odd parity in the WMAP data. In addition, there appears to be a parity-even direction at $90^{\circ}$ from it that does not appear significant.
These results were achieved using a novel masking scheme, that, as far as we can see, is the cleanest way to calculate large scale multipole coefficients from WMAP data.  This masking scheme is a compromise between reducing remaining outliers in the low foreground ILC full sky map and avoiding reconstruction noise and is far from being ideal. It keeps most of the regions masked by the KQ75 mask and it ignores some regions outside the KQ75 mask.

Planck is expected to see better than WMAP though the Galactic noise. Thus the significance of the parity-odd and even directions we found is likely to change. Since we are already reporting at least a 3.6$\sigma$ effect, if the significance increases further, we believe that this will be a real challenge to $\Lambda$CDM, one that calls for either a systematic or cosmological explanation.

One possible cosmological explanation, that in fact motivated us to look for non-trivial parity effects in the CMB, is a moving pre-inflationary particle (PIP). Motivated by \cite{StringTheory} the cosmological imprints of a static PIP were studied in \cite{Sunny,PIP}. The most distinct CMB imprint of a stationary particle was found to be giant concentric rings. Such rings were found in \cite{Rings} with fairly high significance (about 3.1$\sigma$). There is, however, no reason why a PIP should be static in the CMB frame. This motivated us to study the cosmological imprints induced by the velocity of the PIP. As will be reported elsewhere \cite{WorkInProgress}, the most general effect induced by the PIP's velocity is a non-trivial parity structure in the CMB.

Another possibility is a pre-inflationary string. As far as we can see at the moment such an object explains naturally the parity-even direction. However, it is hard to explain the more significant parity-odd direction (and the giant rings of \cite{Rings}) via a pre-inflationary string. It should be interesting to see if other pre-inflationary scenarios, such as bubble collisions (see \cite{Kleban} and references within), could induce the parity effects reported here.

\acknowledgments

We thank the anonymous referee for helpful comments.
We acknowledge the use of the Legacy Archive
for Microwave Background Data Analysis (LAMBDA)
\cite{lambda} and the use of the HEALPix package \cite{healpix}. This work
is supported in part by the Israel Science Foundation
(grant number 1362/08) and by the European Research
Council (grant number 203247).

\end{document}